\documentclass[letterpaper,showpacs,floatfix,aps,prb,amsmath,twocolumn,superscriptaddress]{revtex4}

\usepackage{epsfig}
\usepackage{verbatim}
\usepackage{epsf}
\usepackage{array}
\usepackage{hyperref}

\newcommand{\be}{\begin{equation}}
\newcommand{\ee}{\end{equation}}
\newcommand{\bea}{\begin{eqnarray}}
\newcommand{\eea}{\end{eqnarray}}
\newcommand{\w}{\omega}
\newcommand{\W}{\Omega}

\renewcommand{\Im}{\mathrm{Im}}

\newcommand{\eref}[1]{Eq.~(\ref{#1})}
\newcommand{\reqs}[1]{Eqs.~(\ref{#1})}
\newcommand{\rref}[1]{(\ref{#1})}

\newcommand{\nn}{\boldsymbol{n}}
\newcommand{\rr}{\boldsymbol{r}}
\newcommand{\A}{\boldsymbol{A}}

\newcommand{\ocite}[1]{Ref.~\onlinecite{#1}}

\begin{document}

\title{Temperature dependence of the superheating field for superconductors in the
high-$\kappa$ London limit}

\author{G. Catelani}
\affiliation{Laboratory of Atomic and Solid State Physics, Cornell
University, Ithaca, New York 14853, USA}
\affiliation{Department of Physics and Astronomy, Rutgers University, Piscataway, New Jersey 08854, USA}

\author{James P. Sethna}
\affiliation{Laboratory of Atomic and Solid State Physics, Cornell
University, Ithaca, New York 14853, USA}

\begin{abstract}
We study the metastability of the superheated Meissner state in type II superconductors
with \ $\kappa\gg 1$ beyond Ginzburg-Landau theory, which is applicable
only in the vicinity of the critical temperature. Within Eilenberger's
semiclassical approximation, we use the local electrodynamic response of the
superconductor to derive a generalized thermodynamic potential valid at any temperature.
The stability analysis of this functional yields the temperature dependence of the superheating
field. Finally, we comment on the implications of our results for superconducting
cavities in particle accelerators.
\end{abstract}
\date{\today}
\maketitle

\section{Introduction}
\label{sec:intro}

The Meissner effect -- the expulsion of a weak magnetic field from a bulk superconductor --
is one of the hallmark of superconductivity.
As the field is increased, the response of a superconductor depends on the value of the
Ginzburg-Landau (GL) parameter $\kappa=\lambda/\xi$, where $\lambda$ and $\xi$ are the magnetic field
penetration depth and the superconducting coherence length, respectively.
Type I superconductors ($\kappa < 1/\sqrt{2}$) usually
turn normal at the thermodynamic critical field $H_\mathrm{c}$, but superconductivity can
be maintained, as a metastable state, up to the superheating field $H_\mathrm{sh}>H_\mathrm{c}$.
As for type II superconductors ($\kappa > 1/\sqrt{2}$), above the first critical field $H_{\mathrm{c}1}$
their stable state is characterized by the presence of vortices,
and superconductivity persists up to the second critical field $H_{\mathrm{c}2}$.
However, since the work of Bean and Livingston\cite{BL}
it is known that an energy barrier at the surface impedes the penetration of vortices into
the bulk, making it possible for the Meissner state to exist as a metastable state up to
a superheating field $H_\mathrm{sh}>H_{\mathrm{c}1}$. Therefore $H_\mathrm{sh}$ is a
characteristic property of both type I and II superconductors. In this paper we consider
the temperature dependence of the superheating field in strong Type II superconductors
with  $\kappa \gg 1$ -- i.e., in the London (or local) limit.

Over the years the issue of the stability of the superheated Meissner state has received much
attention. The simplest system in which this problem can be studied is a clean superconductor
occupying a half space with a magnetic field applied parallel to the surface.
For this system in the strong type II limit, and assuming that the instability is due to
fluctuations in the direction perpendicular to the surface (i.e., one-dimensional
fluctuations), de Gennes\cite{deG} calculated the superheating field
$H_\mathrm{sh}=H_\mathrm{c}$ near the critical temperature $T_\mathrm{c}$. If the instability
signals the penetration of vortices, however, the relevant fluctuations can be expected
to vary along two dimensions while preserving translational invariance along the field direction.
Galaiko\cite{gal} showed that this is indeed the case and near $T_\mathrm{c}$ the actual superheating
field is smaller than that found by de Gennes, $H_\mathrm{sh}\simeq 0.745 H_\mathrm{c}$.
More details about the critical fluctuations were presented by
Kramer,\cite{kram,kram2} especially in relation to the problem of vortex nucleation.
The question of metastability has also attracted the interest of the mathematical community,\cite{math}
and a detailed
study of the instability due to one-dimensional fluctuations in type II superconductors
was presented in Ref.~\onlinecite{chap}. A similar analysis was performed
for type I superconductors in Ref.~\onlinecite{DDD}, in which the results of earlier
numerical\cite{msj,fp} and analytical\cite{parr} investigations are confirmed and extended.

It is interesting to note that all the previous calculations of the superheating field were
performed within the GL theory (with the exception of Ref.~\onlinecite{gal} in which the
zero-temperature limit is also considered).
This approach, however, is justified only near the critical temperature
(i.e., if $T_\mathrm{c}-T\ll T_\mathrm{c}$), and
a quantitative evaluation of the superheating field at low temperatures
requires the use of the microscopic theory of superconductivity.
Understanding the temperature dependence of the superheating field is important
for practical applications: the maximum accelerating field of superconducting cavities
used in particle accelerators
is limited by the superheating field\cite{pada} and the optimal operational temperature
lies well below $T_\mathrm{c}$.\cite{proch}

In this work we consider, within the semiclassical approach of Eilenberger,\cite{Eilen}
a clean type II superconductor occupying the
half space $x>0$ in the presence of an external magnetic field $\boldsymbol{H}_a$ parallel to the surface.
We derive an expression for the thermodynamic potential valid at any temperature
for $\kappa \gg 1$, which enables us to calculate the temperature dependence of the
superheating field.
As a result we find that in the limit $\kappa \to +\infty$ the ratio $H_\mathrm{sh}/H_\mathrm{c}$
between superheating and thermodynamic critical fields is a non-monotonic
function of temperature which has a maximum at $T\approx 0.06 T_\mathrm{c}$.

This paper is organized as follows: in Sec.~\ref{sec:st} we briefly review the semiclassical
theory of superconductivity and introduce our notation, while the derivation of the thermodynamic
potential is presented in Sec.~\ref{sec:tp}.
The conditions for the (meta)stability of the Meissner state are discussed in
Sec.~\ref{sec:sc}, and the superheating field is calculated in Sec.~\ref{sec:sh}.
In Sec.~\ref{sec:app} we discuss the implications of our results for accelerator applications
and a brief summary is given in Sec.~\ref{sec:con}.

\section{Semiclassical theory}
\label{sec:st}

In the semiclassical approach to superconductivity the
superconducting system is described by the set of equations, the
so-called Eilenberger equations, which are valid at any temperature\cite{tempnote}
under the assumption that the Fermi wave length $\lambda_\mathrm{F}$
is the smallest length scale characterizing the system. In practice this means that the
semiclassical approximation usually applies to low-$T_\mathrm{c}$
superconductors, in which the zero-temperature coherence length
$\xi_0\gg\lambda_\mathrm{F}$. This condition also implies the
applicability at magnetic fields as high as $H_{\mathrm{c}2}$.
This technique is widely used to study the properties of hybrid
superconducting devices; see, e.g., Ref.~\onlinecite{belzig}.
From now on, we will use units such that Boltzmann constant $k_B=1$ and
the Planck constant $\hbar=1$.

The Eilenberger equations are equations for the anomalous Green's
functions $f(\w_n,\nn,\rr)$ and $\bar{f}(\w_n,\nn,\rr)$, which depend on
the Matsubara frequencies $\w_n = 2\pi T(n+1/2)$, the position $\rr$,
and the unit vector $\nn$ on the Fermi surface,
\be\begin{split}\label{feq}
\left\{\w_n + \nn\cdot\left[\boldsymbol{\nabla} -
i\boldsymbol{A}(\rr)\right]\right\} f(\w_n,\nn,\rr)
&=\Delta(\rr) g(\w_n,\nn,\rr) \, ,  \\
\left\{\w_n - \nn\cdot\left[\boldsymbol{\nabla} +
i\boldsymbol{A}(\rr)\right]\right\} \bar{f}(\w_n,\nn,\rr)
&=\Delta^\dag(\rr) g(\w_n,\nn,\rr) \, ,
\end{split}\ee
where the dagger denotes complex conjugation.
The (normal) Green's function $g(\w_n,\nn,\rr)$ is related to
$f$ and $\bar{f}$ via the constraint (suppressing all the arguments for brevity):
\be\label{cons}
g^2 + f\bar{f} = 1 \, .
\ee
These equations are to be solved together with the self-consistent
equation for the complex order parameter $\Delta(\rr)$ and the Maxwell equation
relating the magnetic field to the (super)current:
\be\label{sce}
\Delta(\rr) \log\frac{T}{T_\mathrm{c}} +2\pi T
\sum_n\!\left[\frac{\Delta(\rr)}{\w_n}-\int\!\frac{d\nn}{4\pi} f(\w_n,\nn,\rr) \right]\!=0 \, ,
\ee
\be\label{me}
\boldsymbol{\nabla}\times\boldsymbol{H}+i
\frac{1}{\kappa_0^2}
2\pi T \sum_n \int\!\frac{d\nn}{4\pi} \, 3\nn g(\w_n,\nn,\rr) = 0 \, ,
\ee
with $\boldsymbol{H} = \boldsymbol{\nabla}\times\boldsymbol{A}$.

In \reqs{feq}-\rref{me} we used as the unit of length the zero-temperature BCS coherence length
\be
\xi_0 = \frac{v_F}{2\Delta_0} \ ,
\ee
where $v_F$ is the Fermi velocity and $\Delta_0$ is the zero-temperature zero-field order
parameter, which gives the energy unit. The vector potential $\boldsymbol{A}$
is rescaled by $\phi_0/2\pi\xi_0$ and the magnetic field $\boldsymbol{H}$ by
$\phi_0/2\pi\xi_0^2$, with $\phi_0=\pi c/e$ as the flux quantum. These choices of units render
all quantities dimensionless; for example, the BCS critical temperature is
$T_\mathrm{c} = e^{\gamma_E}/\pi\simeq 0.567$, where $\gamma_E$ is Euler's constant.
Finally, the dimensionless
parameter $\kappa_0$, the only independent parameter remaining after the units are chosen,
is defined in analogy with the GL parameter $\kappa$ as
\be
\kappa_0 = \frac{\lambda_0}{\xi_0} \, ,
\ee
where the zero-temperature penetration depth is
\be
\frac{1}{\lambda_0^2}=\frac{8\pi}{3} \left(\frac{2\pi\xi_0}{\phi_0}\right)^2 \nu \Delta_0^2
\ee
with $\nu$ as the density of states at the Fermi energy.

In writing \eref{feq} higher-order terms in the magnetic field are neglected
which give rise to diamagnetic effects.\cite{wert} This is a good approximation if
\be
\w_c \ll T \, ,
\ee
where $\w_c=eH/mc$ is the cyclotron frequency. This condition can be rewritten as:
\be\label{tlim}
\frac{\lambda_F}{\xi_0} \frac{1}{\kappa_0} \frac{H}{H_\mathrm{c}(0)} \ll \frac{T}{T_\mathrm{c}}
\ee
In low $T_\mathrm{c}$, strong type II superconductors, the first two factors on the left-hand side are
both small parameters.
In what follows we consider magnetic fields $H$ smaller than the zero temperature critical field
$H_\mathrm{c}(0)$; therefore, this approximation is justified down to very low temperatures.\cite{tempnote2}

\section{Thermodynamic potential}
\label{sec:tp}

Equations \rref{feq}, \rref{sce}, and \rref{me} are the Euler-Lagrange equations obtained by varying
the following functional\cite{Eilen} with respect to $\bar{f}$ and $f$, $\Delta^\dag$, and
$\boldsymbol{A}$, respectively:
\be\label{tpstart}\begin{split}
\Omega  = & \, \nu\!\int\!d^3r \bigg\{ \frac{\kappa_0^2}{3}
\left(\boldsymbol{H}(\rr)-\boldsymbol{H}_a\right)^2
+ |\Delta(\rr)|^2 \log \left(\frac{T}{T_\mathrm{c}}\right)
\\ & + \int\!(dn) \bigg[ \frac{|\Delta(\rr)|^2}{\w_n}
-\Delta^\dag(\rr) f - \bar{f} \Delta(\rr) \\ & -2\w_n (g-1) - g
\nn\cdot\left(\boldsymbol{\nabla}\log \frac{f}{\bar{f}}-
2i \boldsymbol{A}(\rr) \right)\!\bigg]\!\bigg\},
\end{split} \ee
with $g$ as implicitly defined by \eref{cons}, $\boldsymbol{H}_a$ as the applied field which we assume
uniform, and
\be\label{intdef}
\int\!(dn) \equiv 2\pi T\sum_n  \int\!\frac{d\nn}{4\pi} \, .
\ee
The functional $\Omega$ in \eref{tpstart} is not the thermodynamic potential;
however, for any given $\Delta(\rr)$ and $\boldsymbol{A}(\rr)$ \eref{tpstart} gives the difference between the
potentials in the superconducting and normal states once the solutions to \eref{feq}
for $f$ and $\bar{f}$ are substituted into it.

As a first step in solving \eref{feq}, we recall that the order parameter
can be assumed as real; more precisely, the (gradient of the) phase of the order parameter
can be collected together with the vector potential into a gauge invariant quantity.
All other quantities become gauge invariant as well, and
the new vector potential is proportional to the supercurrent velocity. This
means that in the Meissner state $\boldsymbol{A}$ must vanish deep into the superconductor; the same
holds for its component perpendicular to the surface, as no current leaves the
superconductor. Moreover, $\boldsymbol{H}=\boldsymbol{H}_a$ at the surface.
These arguments determine the boundary conditions
for $\boldsymbol{A}$; further boundary conditions are discussed at the end of this section.

With a real order parameter, it is convenient to introduce the sum and difference of the anomalous
Green's functions,
\be\label{sd}
s = f+ \bar{f}\, , \quad d = f-\bar{f} \, .
\ee
Then, \eref{cons} can be rewritten as
\be\label{cons2}
g^2 = 1 - \frac{1}{4} \left(s^2-d^2\right) \, .
\ee
Taking the difference between the equations in \eref{feq} and solving for $d$ in terms of $s$, we obtain
\be\label{d-eq}
d= - \frac{\nn\cdot\boldsymbol{\nabla}s}{\W_n} \, ,
\ee
where we introduced the short hand notation
\be\label{On-def}
\W_n = \w_n -  i \nn\cdot\boldsymbol{A} \, .
\ee

Our discussion so far is valid for any $\kappa_0$, but to find an explicit expression for the
thermodynamic potential as a functional of  $\Delta(\rr)$ and $\boldsymbol{A}(\rr)$, we look
for an approximate solution to \eref{feq} for $\kappa_0 \gg 1$ and at arbitrary temperature.
To find a suitable approximate expression for $s$, we note that a rescaling
$\rr \to \kappa_0 \rr$ only affects the gradient terms in \eref{feq}, so that an expansion in the
small parameter $1/\kappa_0$ is equivalent to a gradient expansion. To zero order we neglect
the gradient terms, drop $d$ in \eref{cons2}, and using the sum of \eref{feq} arrive at
\be\label{sg0}\begin{split}
s^{(0)} & = \frac{2\Delta}{\sqrt{\W_n^2+\Delta^2}} \, , \\
g^{(0)} & = \frac{\W_n}{\sqrt{\W_n^2+\Delta^2}} \, ,
\end{split}\ee
which for $\boldsymbol{A} =0$ correctly reduce to the standard result\cite{belzig} for a bulk
superconductor in the absence of magnetic field.
Hereinafter, due to the above-mentioned rescaling, $\lambda_0$ is the unit of length.

To calculate the next order in the expansion, we define [see \eref{d-eq}]
\be\label{d1}
d^{(1)}  = - \frac{1}{\kappa_0}\frac{\nn\cdot\boldsymbol{\nabla}s^{(0)}}{\W_n} \, .
\ee
From \eref{cons2} we obtain
\be\label{g2}\begin{split}
g \ \simeq & \, g^{(0)} + g^{(2)} \, ,\\
g^{(2)} = & \, \frac{\left(d^{(1)}\right)^2}{8g^{(0)}} - \frac{s^{(0)}}{4g^{(0)}}s^{(2)} \, ,
\end{split}\ee
where $s^{(2)}$ is the next non-trivial order in the expansion for $s$ (i.e., $s\simeq s^{(0)}+s^{(2)}$),
which is again found from the sum of \eref{feq},
\be\label{s2}\begin{split}
\kappa_0^2 s^{(2)} = & \ \frac{\Delta}{4}\frac{\left(\nn\cdot\boldsymbol{\nabla}s^{(0)}\right)^2}{\Omega_n^2}
\frac{1}{S_n} +\frac{(\nn\cdot\boldsymbol{\nabla})^2 s^{(0)}}{S_n^2}
\\ & +\frac{(\nn\cdot\boldsymbol{\nabla})\hat{A}}{\Omega_n}
\frac{\nn\cdot\boldsymbol{\nabla}s^{(0)}}{S_n^2}
\end{split}\ee
with
\be
S_n = \sqrt{\W_n^2 + \Delta^2} \, .
\ee
It turns out that this expression, however, is not needed to obtain
the thermodynamic potential, as its contributions to it cancel out. This
can be checked by substituting Eqs.~\rref{sd}-\rref{cons2} into \eref{tpstart} and using the approximate
expressions in Eqs.~\rref{sg0}-\rref{g2}. After an integration by parts (and dropping the resulting
surface term), to lowest nontrivial order in the gradient expansion,
the thermodynamic potential as a functional of $\Delta$ and $\boldsymbol{A}$ is
\be\label{tpge}\begin{split}
\Omega = &\,\nu \int\!d^3r \bigg\{\frac{1}{3}\left(\boldsymbol{\nabla}\times\boldsymbol{A}
-\boldsymbol{H}_a\right)^2
+ \Delta^2 \log \left(\frac{T}{T_\mathrm{c}}\right)
\\& + \int\!(dn) \bigg[
\frac{\Delta^2}{\w_n} - 2 \left(\sqrt{\Omega^2_n +\Delta^2} -\w_n \right)
\\ & +\frac{1}{\kappa_0^2}
\frac{\sqrt{\Omega^2_n+\Delta^2}}{4\Omega^2_n}\left(\nn\cdot\boldsymbol{\nabla} s^{(0)}\right)^2
\bigg]\bigg\} .
\end{split}\ee
This expression is one of our main results and is the starting point to study the
metastability of the Meissner state;
see Sec.~\ref{sec:sc}. It can be considered as an extension of the GL approach, and as a check,
we show in Sec.~\ref{sec:GLlim} that \eref{tpge} reduces
to the known GL potential in the appropriate limit. It is interesting to compare
the present result with other extensions of the GL theory in the
literature,\cite{tewo,wert2,kosz,kos,kopi} where the expansion is performed with
respect to the covariant derivative [i.e., the gauge-invariant operator
$\nabla-2ie\A$\cite{footA}].
In the present notation, this amounts to supplementing the already performed gradient expansion with an
expansion over $\A$; at lowest nontrivial order the published results are recovered. As the field
increases, however, $\A$ increases as well and this additional expansion is not reliable. In the local limit
considered here the order parameter amplitude spatial profile is determined primarily by the depairing
effect of the supercurrent, which is taken into account exactly, while the additional gradient
terms are suppressed by the small parameter $1/\kappa_0^2$.\cite{gradexpnote}

\subsection{Ginzburg-Landau limit}
\label{sec:GLlim}

As remarked in the Sec.~\ref{sec:intro}, the GL approach is valid near the critical temperature.
More generally, near a second-order phase transition the order parameter is small and an expansion of the
thermodynamic potential in powers of the small parameter $\Delta/2\pi T$ becomes viable. To perform
this expansion in the present case, we introduce the rescaled vector potential
\be\label{atil}
\tilde{\A} = \sqrt{\frac{2}{3}}\frac{\A}{\Delta(T)} \, .
\ee
Here $\Delta(T)$ is the value of the order parameter at temperature $T$ in zero magnetic field, which
by definition satisfies the equation
\be\label{DTeq}
\log \left(\frac{T}{T_\mathrm{c}}\right)+2\pi T \sum_n \left[ \frac{1}{\w_n} -
\frac{1}{\sqrt{\w_n^2 + \Delta(T)^2}} \right] = 0
\ee
obtained by minimizing \eref{tpge} with $\A=\boldsymbol{H}_a=0$. The
expansion of the above equation near $T_\mathrm{c}$ leads to the well-known\cite{tink} approximate expression
\be
\frac{\Delta(T)}{2\pi T} \approx \sqrt{\frac{2}{\zeta}} \sqrt{1-\frac{T}{T_\mathrm{c}}} \, ,
\ee
where
\be
\zeta = \sum_n \frac{1}{(n+1/2)^3} = -\psi''\left(\frac{1}{2}\right) = 7\zeta(3) \, .
\ee

The rescaling in \eref{atil} is equivalent to normalizing the field with respect to the
temperature-dependent thermodynamic critical field rather than its zero \-temperature value; the
scaling is to be applied to the external field $\boldsymbol{H}_a$ as well. Similarly, we define the
normalized order parameter
\be\label{psi}
\psi(\rr)=\frac{\Delta(\rr)}{\Delta(T)}
\ee
and introduce the temperature-dependent penetration depth
\be\label{lamt}
\lambda(T) = \frac{\lambda_0}{\sqrt{\zeta}}\frac{2\pi T}{\Delta(T)} \approx
\frac{\lambda_0}{\sqrt{2}}\frac{1}{\sqrt{1-T/T_\mathrm{c}}}
\ee
as the unit of length. It should be stressed that all the temperature dependencies
introduced in this subsection are valid only in the vicinity of the
critical temperature; see also Ref.~\onlinecite{tink}.

Substituting the definitions in Eqs.~\rref{atil} and \rref{psi} into \eref{tpge},
expressing lengths via $\lambda(T)$, and expanding in powers
of $\Delta(T)/2\pi T \ll 1$, the lowest-order term is
\be\label{tpgl}\begin{split}
\Omega_{GL}=&\nu\zeta\frac{\Delta^4(T)}{(2\pi T)^2}\int\!d^3r\bigg\{
\frac{1}{2}\left(\boldsymbol{\nabla}\times\tilde{\boldsymbol{A}}-\boldsymbol{H}_a\right)^2
\\ & -\frac{1}{2} \psi^2 + \frac{1}{2} \tilde{A}^2 \psi^2
+ \frac{1}{4} \psi^4 +\frac{1}{2\kappa_{GL}^2}\left(\boldsymbol{\nabla}\psi\right)^2\bigg\}.
\end{split}\ee
The GL parameter $\kappa_{GL}$ is proportional to $\kappa_0$,
\be
\kappa_{GL} = \sqrt{\frac{3}{2\zeta}}\,\kappa_0 \approx 0.42 \kappa_0 \, .
\ee
It can be written as the ratio
\be
\kappa_{GL} = \frac{\lambda(T)}{\xi(T)}
\ee
between the temperature-dependent penetration depth [\eref{lamt}] and coherence length
\be
\xi(T) = \sqrt{\frac{2}{3}}\frac{2\pi T}{\Delta(T)} \xi_0 \, .
\ee

Inside the curly brackets in \eref{tpgl} one can recognize the GL free energy
in its dimensionless form; see, e.g., Refs.~\onlinecite{kram} and \onlinecite{DDD}. The last term,
in particular, represents
the energy associated with the spatial variation of the amplitude of the order parameter.
Due to this term the appropriate boundary condition at the surface is $\psi'=0$, where the prime
indicates the derivative along the normal to the surface. The similar term in \eref{tpge}
has a more complicated dependence on both the order parameter and the vector potential;
for this reason the issue of the boundary condition for $\Delta$ demands further investigation beyond
the scope of the present work; for example,
the addition of surface terms [such as those discarded in obtaining \eref{tpge}] may be necessary;
see, e.g., Ref.~\onlinecite{bcref}.
In what follows, however, we will concentrate on the limit $\kappa_0\to \infty$; in this case the
last term in \eref{tpge} is neglected,
the order parameter is determined by a ``local'' equation (rather than a differential one -- see
Sec.~\ref{sec:sc}), and no
boundary conditions are required for $\Delta(\rr)$.

\section{Stability condition}
\label{sec:sc}

In Sec.~\ref{sec:tp} we have derived an approximate expression [\eref{tpge}] for the thermodynamic
potential valid at large $\kappa_0$. A standard procedure can be now applied to study the properties of
this functional; for example, the self-consistent equation for $\Delta$ and the Maxwell equation
are found by taking the variations with respect to $\Delta$ and $\A$, respectively. Considering
from now on only the lowest order contributions in $1/\kappa_0$
[i.e., neglecting the last term on the right-hand side of
\eref{tpge}] and using \eref{DTeq}, we find
\be\label{d-eq-k}
\int\!(dn)\left[\Delta \left(\frac{1}{\sqrt{\w^2_n+\Delta(T)^2}}
- \frac{1}{\sqrt{\Omega_n^2+\Delta^2}}\right)
\right]\!= 0 \, ,
\ee
which gives, via definition \rref{On-def}, a local relation between order parameter and
vector potential, and
\be\label{a-eq-k}\begin{split}
\boldsymbol{\nabla}\times\boldsymbol{\nabla}\times\boldsymbol{A}
-i \int\!(dn) \, 3\nn
\frac{\Omega_n}
{\sqrt{\Omega_n^2+\Delta^2}} = 0 \, .
\end{split}\ee

Solutions to \reqs{d-eq-k} and \rref{a-eq-k} are (meta)stable only if they are a minimum of $\Omega$; i.e.,
if its second variation is positive. To
investigate the stability, let us parametrize $\Delta$ and $\A$ as
\be
\Delta = \Delta_s + \eta \, , \quad \A = \A_s + \boldsymbol{a} \, ,
\ee
where $\Delta_s$ and $\A_s$ satisfy \reqs{d-eq-k} and \rref{a-eq-k}. Expanding \eref{tpge}
for small $\eta$ and $\boldsymbol{a}$, the second variation is:
\be\label{d2o}\begin{split}
\delta^2\Omega = \nu\int\!d^3r\bigg\{
\int\!(dn) \bigg[
\frac{\Delta_s^2}{(\Omega_s^2+\Delta_s^2)^{3/2}}\left(\eta^2 + (\nn\cdot\boldsymbol{a})^2\right)
\\ -2i\frac{\Omega_s\Delta_s}{(\Omega_s^2+\Delta_s^2)^{3/2}}\eta(\nn\cdot\boldsymbol{a})
\bigg]
+\frac{1}{3}\left(\boldsymbol{\nabla}\times\boldsymbol{a}\right)^2\bigg\}
\end{split}\ee
with $\Omega_s = \w_n - i\nn\cdot\A_s$. In the absence of magnetic field
(i.e., $\A_s=0$) the superconducting state is stable for any $T<T_\mathrm{c}$; after integrating over
$\nn$ the last term in square brackets in \eref{d2o} vanishes, so that
$\delta^2\Omega>0$ for any fluctuation as long as $\Delta_s > 0$.
As the field increases,
however, the sign of $\delta^2\Omega$ changes, by definition, when the superheating field is reached.
Therefore, to find $H_\mathrm{sh}$ we look for nontrivial fluctuations $\eta\neq 0$, $\boldsymbol{a} \neq 0$
such that $\delta^2\Omega[\Delta_s,\A_s] = 0$.

In the geometry under consideration (i.e., superconductor in the $x>0$ half space and
$\boldsymbol{H}_a\parallel z$), the solution to \reqs{d-eq-k} and \rref{a-eq-k} is parametrized
as
\be\label{solform}
\Delta_s = \Delta_s(x) \, , \qquad \A_s = [0,A_y(x),0] \, .
\ee
Then, as shown by Kramer\cite{kram} in the GL limit, the fluctuations can be taken in the following form:
\be\label{fl-form}\begin{split}
\eta & = \tilde{\eta} (x,k) \cos (ky) \, ,\\
\boldsymbol{a} & = [\tilde{a}_x(x,k) \sin(ky) , \tilde{a}_y(x,k) \cos(ky) , 0] \, .
\end{split}\ee
Substituting \eref{fl-form} into \eref{d2o} and minimizing with respect to $\tilde{\eta}$ and
$\tilde{a}_x$, we find
\be
\tilde{\eta} = \frac{G}{F_0} \tilde{a}_y \, ,
\ee
\be
\tilde{a}_x = \frac{k}{3F_x+k^2} \tilde{a}'_y \, ,
\ee
with prime denoting differentiation with respect to $x$,
\be
G =  \int\!(dn) \frac{i\Omega_s \Delta_s n_y}{(\Omega_s^2+\Delta_s^2)^{3/2}}
\ee
and
\be
F_i = \int\!(dn) \frac{\Delta_s^2 n_i^2}{(\Omega_s^2+\Delta_s^2)^{3/2}}.
\ee
Here $n_0 = 1$ and we note the property $|F_i| \leq 1$.
With these definitions, we obtain for the second variation (up to a numerical prefactor)
\be\label{d2ofin}
\delta^2\Omega \propto \int_0^{\infty}\!\!dx \left[
\frac{F_x}{3F_x + k^2}\left(\tilde{a}'_y\right)^2 +
\frac{F_0F_y - G^2}{F_0}\left(\tilde{a}_y\right)^2 \right].
\ee
The first term on the right-hand side of \eref{d2ofin} gives always a positive contribution to
the second variation, but as we now argue it can be neglected in the large $\kappa_0$ limit.
Clearly, the larger $k$ is the smaller this contribution becomes; on the other hand,
the second variation of the last term in \eref{tpge}, which we have neglected, would schematically
contribute a (positive) term proportional to $k^2/\kappa_0^2$. Therefore, the optimal value is
$k \sim \sqrt{\kappa_0}$, which is in agreement with the GL result of \ocite{kram}, and the two terms both
give contributions of $\sim 1/\kappa_0$, which we neglect as $\kappa_0 \to +\infty$. As a consequence,
the superheating field is determined by the vanishing of the coefficient of the second term
on the right-hand side of \eref{d2ofin}, i.e.,
\be\label{stabcon}
F_0F_y - G^2 = 0 \, .
\ee
In Sec.~\ref{sec:sh} we use this condition together with
\reqs{d-eq-k} and \rref{a-eq-k} to calculate the superheating field.

\section{Superheating field}
\label{sec:sh}

The stability analysis of the previous section gives a conceptually simple procedure to
find the superheating field; we should first solve \reqs{d-eq-k} and \rref{a-eq-k}
to find the profile [cf. \eref{solform}] of the (meta)stable superconducting state for a given
applied magnetic field, and then find $H_a$ such that the condition for the instability
threshold in \eref{stabcon} is
satisfied. The task of solving the non-linear differential equation \rref{a-eq-k}, however,
makes this route difficult in practice. An alternative approach is based on the observation that
the combination
\be\label{firstint}
H_s^2 - 3\int\!(dn) \left[
\frac{\Delta^2_s}{\sqrt{\Omega^2_s +\Delta^2_s}} - 2 \left(\sqrt{\Omega^2_s +\Delta^2_s} -\w_n \right)
\right],
\ee
where $H_s = \boldsymbol{\nabla}\times\A_s$, is constant throughout the superconductor,\cite{footcons}
as can be checked using \reqs{d-eq-k} and \rref{a-eq-k}. Taking into account the boundary conditions, the
form of the solution in \eref{solform}, and the expression
\be\begin{split}
H_\mathrm{c}^2(T) = 6\pi T \sum_n\Bigg[&2\left(\sqrt{\w^2_n +\Delta(T)^2} -\w_n \right)
\\ & -\frac{\Delta(T)^2}{\sqrt{\w^2_n +\Delta(T)^2}}\Bigg]
\end{split}\ee
for the critical field, we find (cf. \ocite{gal})
\be\label{haeq}
H_a^2 = H_\mathrm{c}^2 + 6\pi T \sum_n \left[2\w_n
+\frac{1}{A_0} \Im \left(\Omega_0 \sqrt{\Omega_0^2 +\Delta_{s0}^2}\right)\right]
\ee
where $A_0 = A_y(0)$, $\Omega_0 = \w_n - iA_0$, and $\Delta_{s0}=\Delta_s(0)$.
The above equation relates the applied field to the values of the order parameter and vector
potential at the surface. At the superheating field, these two quantities can be found by solving
the local \reqs{d-eq-k} and \rref{stabcon}. This can be done analytically in the limiting cases
$T\to T_\mathrm{c}$ and $T\to 0$, as we now show.

\subsection{Limiting cases}
\label{sec:lc}

In the GL ($T\to T_\mathrm{c}$) limit \reqs{d-eq-k} and \rref{stabcon} reduce to,
respectively,
\be
\Delta_{s0}^2 + \frac{2}{3}A_0^2 = \Delta(T)^2 \, ,
\ee
\be
\frac{1}{3}\Delta_{s0}^4 - \frac{4}{9}A_0^2\Delta_{s0}^2 = 0 \, .
\ee
Solving these equations we find $\Delta_{s0} = \sqrt{2/3} \Delta(T)$. Defining
\be
\tilde{H} \equiv \frac{H_\mathrm{sh}}{H_\mathrm{c}}
\ee
and substituting the result into the GL limit of \eref{haeq}
\be
\tilde{H}^2 = 1 - \frac{\Delta_{s0}^4}{\Delta(T)^4} \, ,
\ee
we arrive at $\tilde{H} = \sqrt{5}/3 \simeq 0.745$, which is in agreement with
Refs.~\onlinecite{gal} and \onlinecite{kram}. This is not surprising, since we showed in Sec.~\ref{sec:GLlim}
the reduction of our thermodynamic potential [\eref{tpge}] to the GL one [\eref{tpgl}] in this limit.

In the opposite limit $T\to 0$ and using the
notation $\lambda=\Delta_{s0}/A_0$, \reqs{d-eq-k} and \rref{stabcon} become\cite{footT0st}
\be\label{0Tsc}
\begin{array}{lcl}
\log \left(\Delta_{s0} \right)= 0 && (A_0<\Delta_{s0}) \, ,\\
\\
\log \left[A_0(1 + \sqrt{1-\lambda^2})\right]&-&\sqrt{1-\lambda^2} = 0
\\ &&  (\Delta_{s0} < A_0 < e/2) \, ,
\end{array}
\ee
\be\label{0Th}\begin{split}
\left[1-\sqrt{1-\lambda^2}\right]\frac{1}{3}\left[1-\sqrt{1-\lambda^2}\left(1+2\lambda^2\right)\right]
\\ - \left[\lambda\sqrt{1-\lambda^2}\right]^2 = 0 \, ,
\end{split}\ee
while \eref{haeq} can be written as
\be
\tilde{H}^2 = 1 - \frac{2}{3} A_0^2 \left[
\frac{1}{2} - \frac{3}{2}\left(1-\lambda^2\right)+\left(1-\lambda^2\right)^{3/2} \right] \, .
\ee
Substitution of the solution to \reqs{0Tsc} and \rref{0Th} into the latter expression finally
gives $\tilde{H} \simeq 0.840$, as found in \ocite{gal}.

\subsection{Temperature dependence}

\begin{figure}
\includegraphics[width=0.47\textwidth]{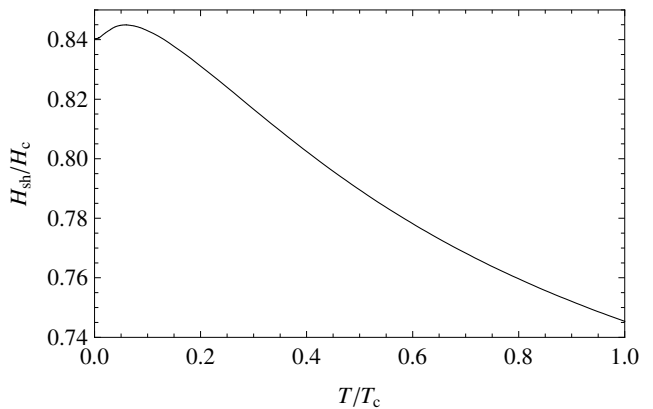}
\caption{Temperature dependence of the ratio $H_\mathrm{sh}/H_\mathrm{c}$. Note the nonmonotonic behavior
with a maximum at low temperature $T \approx 0.06 T_\mathrm{c}$.}
\label{fig}
\end{figure}

\begin{figure}[b]
\includegraphics[width=0.47\textwidth]{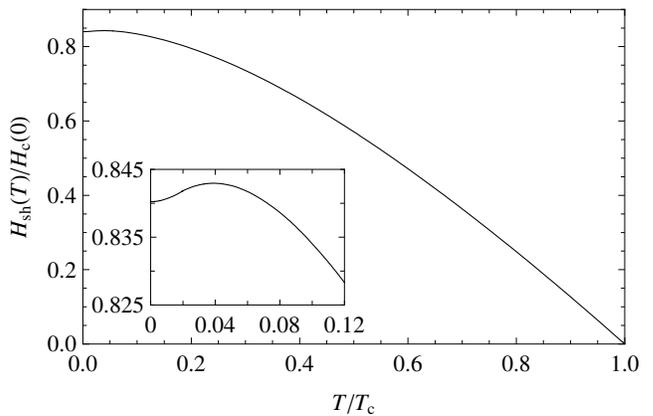}
\caption{Temperature dependence of $H_\mathrm{sh}$ normalized by the
zero-temperature critical field. The nonmonotonic behavior
with a maximum at $T \approx 0.04 T_\mathrm{c}$ is evident in the inset, which
zooms in on the low-temperature region.}
\label{fig2}
\end{figure}

Having verified that the known limiting results are reproduced with our approach, we now consider the
temperature dependence of $\tilde{H}$. Near the two limiting temperatures
we can in principle calculate temperature dependent corrections by further
expanding over the small parameters $\Delta/2\pi T$ close to $T_\mathrm{c}$
and $T/\Delta_0$ as $T\to 0$. Instead, to obtain the behavior at arbitrary temperature,
we resort to a numerical approach. Following the same strategy
as in Sec.~\ref{sec:lc} we first solve numerically the system composed
of \reqs{d-eq-k} and \rref{stabcon} to find $A_0$  and $\Delta_{s0}$; then,  we substitute
the result into \eref{haeq}.
In this way, we obtain the curve presented in Fig.~\ref{fig}. Interestingly, we find a nonmonotonic
dependence of $\tilde{H}$ on temperature, with a maximum of
$\tilde{H}^\mathrm{max} \approx 0.845$ at $T \approx 0.06 T_\mathrm{c}$. A nonmonotonic
behavior is also found in $H_\mathrm{sh}(T)$ shown in Fig.~\ref{fig2}.
Taking into account the decrease in $H_\mathrm{c}(T)$ with increasing temperature,
$H_\mathrm{sh}(T)$ acquires its maximum value
$H_\mathrm{sh}^\mathrm{max} \approx 0.843 H_c(0)$ at the lower temperature $T\approx 0.04 T_\mathrm{c}$,
see the inset of Fig.~\ref{fig2}. Since
$H^2_\mathrm{c}(0) = 4\pi \nu \Delta_0^2$ depends only on material properties, this implies that
in the London limit there is an optimal temperature at which the superheating field is the
highest possible for a given material.

\section{Implications for accelerator design}
\label{sec:app}

The rf cavities used in modern particle accelerators, both
for high-energy physics and as x-ray sources, are made of superconducting
niobium. The best Nb cavities are operated in the metastable region well above $H_\mathrm{c1}$.
The superheating field $H_\mathrm{sh}$ provides an upper bound for the maximum particle
acceleration that a given cavity can produce,\cite{pada}
and the operating point for the best cavities is
approaching the theoretical limit provided
by the GL theory, when the latter is extrapolated to the operating condition
($T\sim 0.2 T_\mathrm{c}$) where it is not valid. Niobium is not a high-$\kappa$ material,
and our theory cannot be directly applied to it, but Fig.\ref{fig} would indicate
that at $T/T_\mathrm{c}\sim 0.2$ the true superheating field would be 11\% higher than the
GL estimate for a high-$\kappa$ material. A change in the theoretical
upper bound of this magnitude would have significant implications for
future attempts to improve the material processing of existing Nb-based
cavities. In principle, a numerical solution of the linear stability
problem for the Eilenberger equations should be possible (albeit challenging)
for all values of $\kappa$, including $\kappa \sim 1$ -- a calculation of direct significance to
current technological applications.

There are several other superconducting materials which appear potentially
promising as eventual replacements for Nb in future accelerator applications, all of
which have significantly higher $\kappa$ and hence are potentially better described
by our London limit calculation. For example,
if run at the current operating temperature of 2~K,
cavities made of Nb$_3$Sn or MgB$_2$ would be near the peak
of $H_\mathrm{sh}/H_\mathrm{c}$ in Fig.~\ref{fig}, and hence our calculation would suggest a peak field
13\% higher than that provided by the GL theory. Using
a current design for the superconducting cavity,
our result for $H_\mathrm{sh}$ suggests a theoretical upper bound for the accelerating field
of 200~MV/m, a factor of 4 larger than the operating fields of the best Nb-based cavities.
Material difficulties have so far kept high-temperature
copper-oxygen-based superconductors from being useful in these applications,
but new high-$T_\mathrm{c}$ materials, e.g., iron pnictides,\cite{feas} may provide more forgiving
material properties.
A quantitative estimate of the superheating field in these materials, however,
may demand calculations that incorporate effects that
go beyond the semiclassical analysis of the
present work. For example, a complete description of superconductivity in MgB$_2$ requires
an Eliashberg-type calculation,\cite{eliashberg} with material properties extracted from a
density-functional electronic structure calculation.\cite{mgb2}

We point out that our result is in sharp contrast with the commonly-used heuristic
$H_\mathrm{sh}\sim H_\mathrm{c}/\kappa$ of Yogi \textit{et al}.\cite{yogi}
This heuristic, termed as the
``line nucleation model'', is not a linear stability calculation, but
an energy balance argument that gives a nonsensical estimate $H_\mathrm{sh} < H_\mathrm{c1}$
for large $\kappa$. The formula's success in describing experiments\cite{yogi}
suggests that there may be nucleation mechanisms (perhaps
disorder-mediated) that become more difficult to control in high-$\kappa$
materials, but it should be viewed as an experimental extrapolation, rather
than a theoretical bound, in guiding the exploration of new materials.

\section{Summary and open problems}
\label{sec:con}

In this paper we have revisited the problem of evaluating the superheating field for type II
superconductors, in particular with regard to its dependence on temperature. To extend previous
calculations\cite{deG,kram,kram2,math} based on the Ginzburg-Landau theory, which is
restricted to temperatures close
to the critical one, we have employed the semiclassical approach in order to derive an approximate
expression for the thermodynamic potential [\eref{tpge}] valid at large values of the Ginzburg-Landau
parameter $\kappa \gg 1$. From this expression we have calculated, in the limit
$\kappa \to +\infty$,  the temperature dependence of the
ratio between the superheating and critical fields which is presented in Fig.~\ref{fig}.
The relevance of our results to applications in
particle accelerators is discussed in Sec.~\ref{sec:app}.

Many natural extensions of this work come to mind,
beyond and in connection with those already mentioned in Sec.~\ref{sec:app}.
For example, it would be interesting
to study in more detail the critical variations, as done in \ocite{kram} in the
GL limit, and the finite $\kappa$ corrections.
In our calculations we have considered the simplest possible case, namely a clean
superconductor with a spherical Fermi surface. However, the semiclassical theory
can easily accommodate anisotropies in the Fermi surface. The effect
of bulk impurities can also be incorporated in the formalism.
In superconducting cavities in the presence of rf fields various mechanisms for the
breakdown of superconductivity are associated with characteristics of the surface, e.g.,
the presence of surface impurities or steps caused by grain boundaries.\cite{pada}
Moreover surface properties, namely specular vs. diffuse reflection, are known to
affect the electromagnetic response of impure superconductors.\cite{halb}
Therefore, it would be important to explore theoretically the impact
of surface imperfections on the superheating field.\cite{brand}

\acknowledgments

We would like to thank Hasan Padamsee for suggesting this project, and
thank him, Matthias Liepe, and Mark Transtrum for helpful and stimulating
conversations. This work was supported by NSF Supplement No. PHY-02020708
and by NSF Grant No. NSF-DMR-0547769.


\begin{thebibliography}{99}
\bibitem{tempnote}See, however, the discussion at the end of this section.
\bibitem{tempnote2}Note that even at $H=H_{\mathrm{c}2}(0)\sim \kappa_0 H_\mathrm{c}(0)$, the
neglected terms are small corrections for low-$T_\mathrm{c}$ materials but
the condition in \eref{tlim} is more restrictive on the temperature at higher fields.
\bibitem{footA}Here the vector potential is the original noninvariant one, without the phase gradient
contribution discussed after \eref{tpstart}.
\bibitem{gradexpnote}This approximation breaks down if the order parameter changes by an amount of
order $\Delta_0$ (i.e., order 1 in our notation) over a length scale comparable to $\xi_0$ ($=1/\kappa_0$
in our units), so that $|\nabla \Delta| \sim k_0$. This
is the case, for example, near a vortex core at low temperatures (but not at sufficiently high temperatures,
since the change in $\Delta$ is approximately limited by $\Delta(T) \le \Delta_0$ over a length scale of order
$\xi(T) \ge \xi_0$) or for nonuniform applied fields with large variations over the same length
scale (i.e., such that $\xi_0|\nabla H|/H \sim 1$).
\bibitem{footcons}This can be justified by treating $x$ as a time coordinate, $\A$ and $\Delta$
as generalized coordinates, and $\Omega$ as an action; then \eref{firstint} corresponds to
the energy conservation law.
\bibitem{footT0st}We note that \eref{0Th} is valid when $A_0> \Delta_{s0}$ (i.e., $\lambda < 1$); if
$A_0 < \Delta_{s0}$, then the right-hand side of \eref{stabcon} is simply $1/3$ and the equality
is not satisfied.
\end{thebibliography}
\end{document}